\documentclass{JHEP3}

\usepackage{graphicx,psfrag,verbatim}
\usepackage{graphicx,verbatim}
\usepackage{latexsym}
\usepackage{amsmath}
\usepackage{amsfonts}
\usepackage{amssymb}

\newcommand{\R}{\mathbb{R}}

\newcommand{\comm}[2]{\left[ #1 , #2 \right]}

\newcommand{\be}{\begin{equation}}
\newcommand{\ee}{\end{equation}}
\newcommand{\bea}{\begin{eqnarray}}
\newcommand{\eea}{\end{eqnarray}}

\newcommand{\ben}{\begin{eqnarray}}
\newcommand{\een}{\end{eqnarray}}
\newcommand{\bes}{\begin{subequations}}
\newcommand{\ees}{\end{subequations}}

\title{Noncommutative field gas driven inflation}

\author{Luciano Barosi $^{a}$, Francisco A. Brito $^{a}$ and Amilcar R. Queiroz $^{b}$ \\
$^{a}$Departamento de F\'\i sica, Universidade Federal de Campina
Grande, Caixa Postal 10071, 58109-970  Campina Grande, Para\'\i ba,
Brazil\\
$^{b}$Centro Internacional de F\'\i sica da Mat\'eria Condensada,
Universidade de Bras\'\i lia, Caixa Postal 04667, Bras\'\i lia, DF,
Brazil\\
E-mail:{ lbarosi@ufcg.edu.br, fabrito@df.ufcg.edu.br,
amilcarq@gmail.com}}

\abstract{We investigate early time inflationary scenarios in an
Universe filled with a dilute noncommutative bosonic gas at high
temperature. A noncommutative bosonic gas is a gas composed of
bosonic scalar field with noncommutative field space on a
commutative spacetime. Such noncommutative field theories was
recently introduced as a generalization of quantum mechanics on a
noncommutative spacetime. As key features of these theories are
Lorentz invariance violation and CPT violation. In the present study
we use a noncommutative bosonic field theory that besides the
noncommutative parameter $\theta$ shows up a further parameter
$\sigma$. This parameter $\sigma$ controls the range of the
noncommutativity and acts as a regulator for the theory. Both
parameters play a key role in the modified dispersion relations of
the noncommutative bosonic field, leading to possible striking
consequences for phenomenology. In this work we obtain an equation
of state $p=\omega(\sigma,\theta;\beta)\rho$ for the noncommutative
bosonic gas relating pressure $p$ and energy density $\rho$, in the
limit of high temperature. We analyse possible behaviours for this
gas parameters $\sigma$, $\theta$ and $\beta$, so that
$-1\leq\omega<-1/3$, which is the region where the Universe enters
an accelerated phase.
 \\

\vspace{1cm}

Keywords: Noncommutative Fields, Lorentz Invariance Violation, CPT Violations,
Inflationary Cosmology}

\maketitle

\begin{document}

\section{Introduction}

Quantum Field Theories (QFT) with Lorentz invariance violation are
subject of growing attention among theoretical physicists in the
last few years
\cite{Kostelecky:1988zi,Kostelecky:1995qk,Colladay:1998fq,Kostelecky:1999mr,Jackiw:1999yp,Kostelecky:2000mm}.
Indeed, there is no reason a priori to believe that the cosmological
principle\footnote{The cosmological principle states that spacetime
is locally Lorentz invariant.} is true at all energy scales. Further
analysis on the problem of matter-antimatter asymmetry
\cite{Dine:2003ax}, ultra high energy cosmic rays
\cite{Anchordoqui:2002hs,Takeda:2002at,Abraham:2006ar}, primordial
magnetic field \cite{Grasso:2000wj}, neutrino physics
\cite{Athanassopoulos:1997pv} and some other cosmological
measurements demands further criticisms on the cosmology principle
at ultra high energy scale. Therefore the proposal of validity/bound
tests for the cosmology principle, i.e., cosmological phenomenology
based upon QFT with Lorentz invariance violation is a subject of
major relevance.

There are distinct approaches to QFT with Lorentz invariance
violation. A first approach is based on the noncommutativity of
spacetime itself. The most studied situation was that of QFTs on the
Groenenwold-Moyal \cite{Szabo:2001kg,Douglas:2001ba}. Some of these
models may be obtained from a fundamental theory such as
string/M-theory \cite{Douglas:2001ba}. Recently
\cite{Chaichian:2004za,Aschieri:2005yw,Aschieri:2005zs}, it was
shown that one could restore Lorentz invariance of these QFT by a
proper twisting of the action of the symmetry group on the fields.
Cosmology applications of this twisted Lorentz invariant QFT is
recently being under study \cite{Akofor:2007fv}. There is also an interesting approach based on using a noncommutative version of Wheeler-deWitt equation, where the noncommutativity is introduced as a Moyal deformation \cite{GarciaCompean:2001wy}.

A second approach is known as extended standard model
\cite{Colladay:1998fq}. It is based on a famous work of S. Carroll,
R. Jackiw and G. Field \cite{Carroll:1989vb}. In this approach the
breaking of Lorentz invariance is performed by the introduction of a
Chern-Simons term on the action of some gauge theory\footnote{It is
worth mentioning that there is some controversy on the correctness
of this approach \cite{Bonneau:2006ma}}. One should also mention the
approach known as doubly special relativity \cite{Magueijo:2001cr}.

Another approach, the one we are going to consider in this paper, is
QFT based on noncommutative fields \cite{Carmona:2002iv,Carmona:2003kh,Balachandran:2007ua}.
On such theories one breaks
Lorentz invariance by modifying the canonical commutation relations
of the fields (therefore the name noncommutative fields) by the
inclusion of an ultraviolet or/and infrared scale factor. These
noncommutative fields arise as a generalization of the quantum
mechanics on the Groenenwold-Moyal plane \cite{Nair:2000ii}, where
one considers the noncommutativity in the degrees of freedom of the
fields. One interesting feature of such theories is that there is a
connection of this approach with Carroll et al proposal
\cite{Carroll:1989vb} for gauge theories. See for instance
\cite{Carmona:2002iv,Gamboa:2005bf,Falomir:2005it,Gamboa:2005pd,Falomir:2006hp}.

One common feature of all these theories with Lorentz invariance
broken is the deformation of the dispersion relations of the plane
waves. It is this deformation that allows one to build
phenomenological models that could be in principle be tested in
future cosmology or particle physics experiments. The kind of
deformation to be seen or ruled out by these experiments will thus
decide which theory will survive.

The most promising arena for these phenomenological models is the
inflationary scenario of the Universe. Models based on the Universe
inflation has successfully explained several observational data,
such as the origin of density fluctuations on large scales. The
general belief is that inflation that supports large and fast growth
of scales can connect the region where the Lorentz invariance might break (i.e., at very high energy)
to cosmological scales. On the other hand, it is known that so far
there is no explanation of the essential mechanisms of inflation
based on fundamental physics such as string/M-theory.


In the context of inflationary scenarios based on QFT with Lorentz
invariance violation, based on the noncommutativity of space-time,
several developments have been put forward recently
\cite{Alexander:2001dr,Lizzi:2002ib,Hassan:2002qk,Koh:2007rx,Koh:2007wa,Chu:2000ww,
Bertolami:2002eq}.
In these works several efforts have been made in the direction of
finding imprints of primordial fluctuations that can survive the
inflationary phase. Such imprints have been investigated both in
theories with \cite{Chu:2000ww} and without \cite{Koh:2007rx}
inflaton fields. See \cite{Brandenberger:2007rg} and references
therein for a recent review on this subject.

In the present paper, we consider the approach based on
noncommutative fields as studied by A. P. Balachandran et al. in
\cite{Balachandran:2007ua}. In that work a noncommutative free
massless bosonic field was studied. For that theory, they considered
a regularization of the delta function appearing in one of the
commutators of the fields. The equal time commutation relation for
the fields then reads
\begin{equation}
\label{eq:comm-rel-modified}
 \left[ \hat{\varphi}^a(x;t),\hat{\varphi}^b(y;t)\right]=i\epsilon^{ab}\theta(\sigma;x-y),
\end{equation}
where $\theta(\sigma;x-y)$ is a Gaussian distribution with a
parameter $\sigma$ related with the standard deviation of the
distribution. Besides introducing one further parameter $\sigma$,
the resulting theory is finite. That fact allowed the study of the
deformed black body radiation spectrum. One interesting feature of
this deformed spectrum is that the energy density for boson with
higher frequency is greater than that for the usual black body
radiation spectrum.

We investigate here an inflationary model based on an Universe filled
with a similar gas composed of a noncommutative boson in the
relativistic limit. In this scenario we study the thermodynamics of
such gas. We aim at computing and analyzing the equation of state
$p=\omega\rho$, with $p$ being the pressure, $\rho$ being the energy
density. In the present situation this equation of state is far from
trivial. Therefore we have to make some careful approximations, such
as considering a power series of the parameters $\theta$ and
temperature $1/\beta$. Now, for suitable values of the parameters
$\theta$, $\sigma$ and the temperature $1/\beta$ of the gas, one obtain $-1\leq\omega< -1/3$. For these
values of $\omega$ the Universe presents an accelerated phase.
Furthermore, the present case should be contrasted with the tachyon
and Chaplygin cosmology that is governed by an equation of state
given by the general form $p=-A/\rho$ \cite{Gibbons:2003gb}, since
here the inflationary regime is followed by a radiation
dominated epoch for sufficiently small density $\rho$.

The present paper is organized as follows: in Sec.~\ref{prel} we
review the formulation of the free noncommutative boson field as
presented in Balachandran et al. The goal is to obtain the
Hamiltonian for such noncommutative field so that we may formulate a
statistical quantum mechanics problem for the gas at high
temperature; in Sec.~\ref{stme} we study the gas of noncommutative
bosons in the grand canonical formalism. Then by some suitable
approximations, we obtain an equation of state
$p=\omega(\sigma,\theta;\beta)\rho$ for this gas; in
Sec.~\ref{cosmonq} we consider a cosmology model with the
noncommutative gas by presenting the proper Friedman equation and
then the behavior of the energy density of this gas with the scale
parameter. Next we consider, by fine tuning the parameters, possible
scenarios yielding inflation; finally in Sec.~\ref{conclu} we finish
the paper with some concluding remarks.

\section{QFT with Noncommutative Target Space  }
\label{prel}

In this section we review the theory \cite{Balachandran:2007ua} of massless spinless
noncommutative bosonic scalar field on a $(3+1)$-d commutative base
space. We consider a commutative $\R^2$ as
target space so that
\begin{equation}\label{eqNCTA:61}
\begin{array}{cccl}
\varphi:&M_3\times \R &\longrightarrow& \R^2 \\
        &(\vec{x},t)         &\longmapsto    & \varphi(\vec{x},t).
\end{array}
\end{equation}
In the sequel we will assume that each spatial direction is compactified into $S^1$ with radius $R$.
Besides making the analysis easier,
this choice of topology of the space will be immaterial, since we
will be interested in the thermodynamic limit, so that $V\sim
R^3\to\infty$.

We write the field components $\varphi^i(\vec{x},t)$ in Fourier
modes as
\begin{equation}
  \label{eqNCTA:62}
  \varphi^i(\vec{x},t) = \sum_{\vec{n}} \;e^{\frac{2\pi i}{R}\vec{n}\cdot \vec{x}} \; \varphi^i_{\vec{n}}(t)\;,
\end{equation}
where $\vec{n}=(n_1,n_2,n_3)$, $n_i\in\mathbb{Z}$, and the Fourier
components are
\begin{equation}
  \label{eqNCTA:63}
  \varphi^i_{\vec{n}}(t) = \frac{1}{R^3} \int d^3x \; e^{-\frac{2\pi i}{R}\vec{n}\cdot \vec{x}} \; \varphi^i(x,t)\;.
\end{equation}

The Lagrangian for the massless spinless boson is given by
\begin{equation}
  \label{eqNCTA:64}
L = \frac{g}{2} \sum_i \int d^3x \big[ \big(\partial_t \varphi^i
\big)^2 - \big(
  \nabla \varphi^i \big)^2 \big],
\end{equation}
which is written in Fourier modes of $\varphi^i$ (\ref{eqNCTA:63})
as
\begin{equation}\label{eqNCTA:65}
  L =  \frac{gR^3}{2} \sum_{i} \bigg\{\dot{\varphi}^i_{\vec{n}} \dot{\varphi}^i_{-\vec{n}} -
  \bigg( \frac{2\pi |\vec{n}|}{R}\bigg)^2 \varphi^i_{\vec{n}} \varphi^i_{-\vec{n}} \bigg\}\;.
\end{equation}
The canonical momenta associated with the Fourier modes
$\dot{\varphi}^i_{\vec{n}}$ are
\begin{equation}
  \label{eqNCTA:66}
  \pi^i_{\vec{n}} = \frac{\partial L}{\partial \dot{\varphi}^i_{\vec{n}}} = gR^3\,\dot{\varphi}^i_{-\vec{n}}.
\end{equation}

In \cite{Balachandran:2007ua}, the scalar field with noncommutative
$\R^2$ as target space denoted by $\hat{\varphi}^a(\vec{x},t)$ was
written in terms of dressed transformations as
\begin{eqnarray}
\hat{\varphi}^{a}(\vec{x},t)&=&\varphi^{a}(\vec{x},t)-
\frac{1}{2}\epsilon^{ab}\theta \pi_{b}(\vec{x},t) \label{eqNCTA:67} \\
\hat{\pi}_{a}(\vec{x},t)&=&\pi_{a}(\vec{x},t)\;, \label{eqNCTA:68}
\end{eqnarray}
where $\theta$ is known as the noncommutative parameter. This
parameter $\theta$ has dimension of (length)$^{-2}$.

The commutation relations at equal time for the noncommutative
scalar fields introduced above are
\begin{eqnarray}
\comm{\hat{\varphi}^{a}(\vec{x},t)}{\hat{\varphi}^{b}(\vec{y},t)}&=&i
\epsilon^{ab}\theta\delta(\vec{x}-\vec{y}), \label{eqNCTA:69}\\
\comm{\hat{\pi}_{a}(\vec{x},t)}{\hat{\pi}_{b}(\vec{y},t)}&=&0, \label{eqNCTA:69b}\\
\comm{\hat{\varphi}^{a}(\vec{x},t)}{\hat{\pi}_{b}(\vec{y},t)}&=&i\,\delta^a_b\delta(\vec{x}-\vec{y})\;.\label{eqNCTA:69c}
\end{eqnarray}
Observe that the usual commutative theory is obtained via $\theta\to
0$.

The novelty of the work of \cite{Balachandran:2007ua} was to
regularise the delta function in the commutation relation
(\ref{eqNCTA:69}). This proved to be useful in the study of the
spectrum of the deformed black-body radiation, where it prevented
the energy density to diverge with respect to frequency. Thus, in
(\ref{eqNCTA:69}) we replace $\theta(\sigma)\equiv\theta(\sigma;x-y)=\theta\delta(x-y)$
by
\begin{equation}\label{eqNCTA:70}
\theta(\sigma)=\frac{\theta}{(\sqrt{2\pi}\sigma)^3} \;\exp \left[-
\sum_{i=1}^3 \frac{(x_i-y_i)^2}{2\sigma^2} \right],
\end{equation}
where we make the simplifying assumption
$\sigma_1=\sigma_2=\sigma_3=\sigma$ in the more general expression
$\sum_{i=1}^3\frac{(x_i-y_i)^2}{2\sigma_i^2}$ for the argument of
the exponential. The delta function in commutation relation
(\ref{eqNCTA:69c}) is not regularized.

We now write the noncommutative fields
$\hat{\varphi}^{a}(\vec{x},t)$ in Fourier modes as
(\ref{eqNCTA:62}), i.e.,
\begin{equation} \label{eqNCTA:72}
\hat{\varphi}^{a}(\vec{x},t)=\sum_{\vec{n}}e^{\frac{2\pi
i}{R}\vec{n}\cdot\vec{x}}\hat{\varphi}_{\vec{n}}^{a}.
\end{equation}
Furthermore, $\theta(\sigma)$ may be written as
\begin{equation}
   \label{eqNCTA:73}
  \theta(\vec{n})=\theta\; e^{-\frac{2\pi^2\sigma^2 |\vec{n}|^2}{R^2}},
\end{equation}
so that the Dressed transformation are now written as
\begin{eqnarray}
\hat{\varphi}_{n}^{a}&=&\varphi_{n}^{a}-\frac{1}{2R^3}\epsilon^{ab}\theta(n)\pi_{-n}^{b}, \label{eqNCTA:74}\\
\hat{\pi}_n^{a}&=&\pi_n^{a}\;. \label{eqNCTA:75}
\end{eqnarray}

The commutation relations for the Fourier components are
\begin{eqnarray}
 \comm{\hat{\varphi}_n^{a}}{\hat{\varphi}_m^{b}}&=&\frac{i\epsilon^{ab}\theta(n)}{R^3}\delta_{n+m,0}\\   \label{eqNCTA:76}
\comm{\hat{\pi}_n^{a}}{\hat{\pi}_m^{b}}&=&0 \label{eqNCTA:77} \\
\comm{\hat{\varphi}_n^{a}}{\hat{\pi}_m^{b}}&=&i\,\delta^{ab}
\delta_{mn} \label{eqNCTA:78}
\end{eqnarray}
and the Hamiltonian reads

\begin{eqnarray}\label{eqNCTA:79}
  H &=& H_0+\sum_{i,\vec{n}} \left(\frac{\Omega^2_{\vec{n}}}{2gR^3} \;
  \pi_{\vec{n}}^i\,\pi_{-\vec{n}}^i +
  \frac{g R^3}{2}\;\omega_{\vec{n}}^2 \,\varphi^i_{\vec{n}} \,
  \varphi^i_{-\vec{n}} -
  \frac{g}{2}\;\omega^2_{\vec{n}}\;\theta(n)\;\epsilon_{ik}\,\varphi^i_{\vec{n}}\,\pi^k_{\vec{n}}\right) \nonumber \\ &&
\end{eqnarray}
where
\begin{equation}\label{eqNCTA:80}
   \Omega^2_{\vec{n}} = 1+\left(\frac{\pi g |\vec{n}|\theta(n)}{R}\right)^2
\end{equation}
and
\begin{equation}\label{eqNCTA:81}
\omega_{\vec{n}} = \frac{2 \pi |\vec{n}|}{R}\;.
\end{equation}

This Hamiltonian may be diagonalized (see section 2 in \cite{Balachandran:2007ua}) by introducing annihilation and
creation operators $A^i_{\vec{n}}$'s and
$A^{i\;\dagger}_{\vec{n}}$'s with $i=1,2$, so that the Hamiltonian
may be written as
 \begin{equation}
  H = \sum_{\vec{n} \neq 0} \omega_{\vec{n}}
\bigg\{ \Lambda^1_{\vec{n}} \, {A_{\vec{n}}^1}^\dagger A_{\vec{n}}^1
+ \Lambda^2_{\vec{n}} \, {A_{\vec{n}}^2}^\dagger A_{\vec{n}}^2
\bigg\}\label{eqNCTA:82}
\end{equation}
with
\begin{eqnarray}
  \Lambda_{\vec{n}}^1 &=&  \Omega_{\vec{n}}  + \frac{\pi|\vec{n}| g \theta(n)}{R} \label{eqNCTA:83}\\
  \Lambda_{\vec{n}}^2 &=&  \Omega_{\vec{n}}  - \frac{\pi|\vec{n}| g \theta(n)}{R}\;. \label{eqNCTA:84}
\end{eqnarray}

We observe that the dispersion relation of these bosons have been
modified. However, if we take the limit $\theta\to 0$, then we
recover the usual dispersion relations. Furthermore, there appears a
splitting of the energy levels of the oscillators. This is
associated with the property of birefringence of the radiation
field.

\section{Noncommutative Field Gas}
\label{stme}
In this section, we formulate the quantum statistical mechanics
problem for the noncommutative field gas in a volume $V\sim R^3$, in
order to study its thermodynamics at some approximation level. The
gas is regarded as a free relativistic ($E\simeq p$) boson gas that
may have dominated the evolution of the hot early Universe during
some time. We use the Hamiltonian defined in Eq.~(\ref{eqNCTA:82})
into the grand partition function determined as
\ben\label{grand}\Xi={\rm
Tr}{e^{-\beta(H-\mu)}}=\prod_{\vec{k}\neq0}\sum_{m,n=0}^\infty
{e^{-\beta(\omega_{\vec{k}}\Lambda_{\vec{k}}^1-\mu)n}e^{-\beta(\omega_{\vec{k}}\Lambda_{\vec{k}}^2-\mu)m}}
\nonumber\\
=\prod_{\vec{k}\neq0}\left(\frac{1}{1-ze^{-\beta\omega_{\vec{k}}\Lambda_{\vec{k}}^1}}
\right)\left(\frac{1}{1-ze^{-\beta\omega_{\vec{k}}\Lambda_{\vec{k}}^2}}
\right),\een where we have omitted the summation on the zero mode
since it is related to the translation of the system. Note that
$z=e^{\beta\mu}$ is the `fugacity', with $\mu$ being a constant chemical
potential.

In the rest of the paper we shall mainly use thermodynamics
quantities that can be found through the following useful function
\ben\label{log_grand}\ln{\Xi}=-
\sum_{\vec{k}\neq0}[\,{\ln{(1-ze^{-\beta\omega_{\vec{k}}\Lambda_{\vec{k}}^1})}
+\ln{(1-ze^{-\beta\omega_{\vec{k}}\Lambda_{\vec{k}}^2})}]}.\een

The relevant quantities of our present investigations are the
internal energy $U$, the thermodynamic number $N$ and pressure $p$
given by
\ben\label{therm_quantity}U=-\frac{\partial}{\partial\beta}\ln{\Xi}(\beta,V,z),\\
\label{therm_quantity2}p=\frac{1}{\beta V}\ln{\Xi}(\beta,V,z), \\
\label{therm_quantity3}N=z\frac{\partial}{\partial z}\ln{\Xi}(\beta,V,z),
\een
being the energy density defined as $\rho=U/V.$ The equation of
state $\rho(p)$ and the Friedmann equation describe the evolution of
a gas-filled Universe.

Given the complexity of  the quantum statistical mechanics problem
formulated for the noncommutative field gas, it is hard to tackle
the problem exactly. Thus let us consider the limit of high
temperature of a dilute gas
($z\ll1).$ At this regime we can easily approximate the equation
(\ref{log_grand}) to the simpler equation
\ben\label{log_grand_approx}\ln{\Xi}= z
\sum_{\vec{k}\neq0}\,{{(e^{-\beta\omega_{\vec{k}}\Lambda_{\vec{k}}^1}}
+{e^{-\beta\omega_{\vec{k}}\Lambda_{\vec{k}}^2})}},\een by expanding
 (\ref{log_grand}) in a power series of $z$ and
retaining only the first-order term. 

We are dealing with a noncommutative field gas composed of $N$
harmonic oscillators enclosed in a volume $V\sim R^3$. In the
thermodynamic limit with $V\to\infty\: (R\to\infty)$ 
the summation is
replaced by an integral on the momenta as follows:
$\sum_{\vec{k}\neq0}\to V\int{\frac{d^3k}{(2\pi)^3}}$. This renders
the following result \ben\label{log_grand_int}\ln{\Xi}= zV
\int{\frac{d^3k}{(2\pi)^3}}\,{{(e^{-\beta\omega_{\vec{k}}\Lambda_{\vec{k}}^1}}
+{e^{-\beta\omega_{\vec{k}}\Lambda_{\vec{k}}^2})}}=\frac{zV}{2\pi^2}
\int_{0}^{\infty}{dk k^2}\,{{\Big(e^{-\beta{{k}}\Lambda^1{(k)}}}
+{e^{-\beta{{k}}\Lambda^2{(k)}}\Big)}},\een with
\ben\Lambda^1{(k)}=\sqrt{1+\left(\frac{k\theta(k)}{8\pi}\right)^2}+\frac{k\theta(k)}{8\pi},\\
\Lambda^2{(k)}=\sqrt{1+\left(\frac{k\theta(k)}{8\pi}\right)^2}-\frac{k\theta(k)}{8\pi},
\een where \ben\theta(k)=\theta e^{-\frac{1}{2}\sigma^2k^2}. \een
The integral (\ref{log_grand_int}) is still hard to evaluate because
of the nontrivial dependence of $\Lambda^i({k})$ in ${{k}}$ and the
parameter $\theta$. However, by expanding the integrand in a power
series of $\theta$ and $\beta$, we are able to advance our
investigations with analytical calculations up to some level of
approximation.

\subsection{The Equation of State}

Here we aim at finding equation of state $p=\omega \rho$, from which one can study possible inflationary scenarios. We should first obtain the thermodynamical quantities $\rho$, the energy density, and $p$, the pressure, defined in (\ref{therm_quantity}) and (\ref{therm_quantity2}), respectively. Using the definition (\ref{log_grand_int}) expanded as a power series of the parameters $\theta$ and $\beta$ up to the fourth order we obtain
\begin{eqnarray}
\rho&=& z\,\left( \frac{12}{{\beta }^4} - \frac{3\,{\theta }^4}{262144\,{\pi }^4\,{\sigma }^8} -
    \frac{15\,\beta \,{\theta }^2}{256\,{\pi }^{\frac{3}{2}}\,{\sigma }^7} + \frac{{\theta }^2}{64\,{\pi }^2\,{\sigma }^6} \right), \label{eq:energy-density2}\\
 p&=& z\left[\frac{4}{{\beta }^4} + {\theta }^2\,\left( \frac{15\,\beta }{512\,{\pi }^{\frac{3}{2}}\,{\sigma }^7} -
     \frac{1}{64\,{\pi }^2\,{\sigma }^6} \right) \right]. \label{eq:pressure}
\end{eqnarray}
 Further useful
quantities for the next section are the particle density $n=N/V$,
defined in (\ref{therm_quantity3}), and the molar entropy $s$
written as
\begin{eqnarray}
 n&=& z\left[\frac{4}{{\beta }^3} + {\theta }^2\,\left( \frac{15\,{\beta }^2}{512\,{\pi }^{\frac{3}{2}}\,{\sigma }^7} -
     \frac{\beta }{64\,{\pi }^2\,{\sigma }^6} \right)\right], \label{eq:number-of-part}\\
 s&=&\frac{160\,{\beta }^8\,{\theta }^2 - 105\,{\sqrt{\pi }}\,{\beta }^7\,{\theta }^2\,\sigma  +
    144\,{\beta }^6\,{\theta }^2\,{\sigma }^2 - 30\,{\sqrt{\pi }}\,{\beta }^5\,{\theta }^2\,{\sigma }^3 +
    2^{14}\,{\pi }^2\,{\sigma }^{10}}{30\,{\sqrt{\pi }}\,{\beta }^9\,{\theta }^2\,{\sigma }^3 -
    16\,{\beta }^8\,{\theta }^2\,{\sigma }^4 + 2^{12}\,{\pi }^2\,{\beta }^4\,{\sigma }^{10}}. \label{eq:entropy}
\end{eqnarray}

Now, the quantity $\omega(\sigma,\theta;\beta)=p/\rho$ in this approximation is found to be
\ben\label{omega}\omega(\sigma,\theta;\beta)=\frac{1}{3}+\frac{\beta^4}{3}
\left(\frac{\theta^4}{2^{20}\pi^4\sigma^8}
-\frac{\theta^2}{2^{6}3\pi^2\sigma^6}\right). \een

One should note that for
 sufficiently large $\sigma$, sufficiently small $\theta$ or sufficiently
small $\beta$ one finds $\omega(\sigma,\theta;\beta)=1/3$, which corresponds to
the radiation dominated regime. It is the minus sign appearing in front of the third term in r.h.s. of Eq. (\ref{omega}) that allows one to find $-1\leq\omega<-1/3$, which is a region of accelerated expansion of an Universe filled with this noncommutative gas.

\section{Cosmology with Noncommutative Field Gas}
\label{cosmonq}

The Friedmann equation governing the evolution of a flat ($k=0$)
Universe is given by \ben
\left(\frac{\dot{a}}{a}\right)^2=\frac{8\pi G}{3}\rho,\een where
$\rho$ is the energy density of several species (for instance,
$\rho_\textrm{radiation}$, $\rho_\textrm{matter}$ and so on)
contributing to the energy density of the Universe and $G$ is the
Newton constant. For a noncommutative field gas dominated epoch, the
energy density for this gas, i.e., $\rho_{ncg}$ in the r.h.s. of the
Friedmann equation is the dominant species component. By using the
equation of conservation of energy density
\ben\dot{\rho}_{ncg}=-3\frac{\dot{a}}{a}(\rho_{ncg}+p_{ncg}), \een
we find \ben \rho_{ncg}\propto a^{-3(1+\omega)}. \een The regime
with $\omega\!=\!-1$ clearly produces $\rho_{ncg}\propto const.$,
which means an exponential expansion.

In the sequel we are going to use the construction of section 3
together with the above Friedmann equation with the noncommutative
gas dominant species component in order to consider some possible
inflationary scenarios. Before that let us summarize what are our
ingredients and what are the expected features of an inflationary
scenario in the present set-up.

We consider an Universe filled with a dilute noncommutative gas at
high temperature described by the quantity
$\omega(\sigma,\theta;\beta)$ given by Eq. (\ref{omega}). We observe
that this noncommutative gas is in principle labelled by three
parameters, i.e, $\sigma$, $\theta$ and $\beta$, each of which we
may be taken either fixed or variable. For inflation to happen,
these parameters should be so that $\omega<-1/3$. We also find
interesting to demand the fields not to be tachyonic, so that we are
going to impose $-1\leq\omega$.

The parameter $\theta$ is related to the intensity of the
noncommutativity. The bigger its value, the bigger the
noncommutativity. The parameter $\sigma$ is related to the range of
the noncommutativity. Now, if $\theta\to 0$, the noncommutativity
disappears. Also if $\sigma\to\infty$, the
noncommutativity becomes negligible. One may easily verify these
statements by analysing Eq. (\ref{eq:comm-rel-modified}) together
with Eq. (\ref{eqNCTA:70}). Now, we observe that if either $\theta$
or $\sigma$ are dynamical variables, then we should look after a
dynamical theory to describe these physical objects. Although we
will analyse situations with $\sigma$ and/or $\theta$ varying, we
will not make statements based on a still to be found dynamical
theory for these variables. The parameter $\beta$ is related to the
inverse of the temperature of the noncommutative gas.

The inflationary scenarios to be considered in the present set-up
should have some special features and requirements. A first
requirement is to see whether the gas can enter and/or exit the
inflationary regime $-1\leq\omega<-1/3$. Of course, we expect the
Universe to enter and then to exit this regime, i.e., one wants to
avoid eternal inflation. Next we are interested on how the
temperature will behave during inflation. Furthermore, we surely
expect either the noncommutative gas to be turned off or its
influence on Friedmann equation be negligible at some time after
exiting inflation. For instance, after inflation it would be
desirable to have a hot Universe in radiation dominated phase.

Having these informations, we may proceed to engineering of some inflationary scenarios.

\subsection{Inflationary Scenarios  with Fixed $\theta$}

Here we consider the noncommutativity parameter fixed, inherently
tied to the short distance physics at all times and consider the
possible inflationary scenarios, by constraining $\beta$ and
$\sigma$. In these scenarios, the classical radiation dominated era
may be achieved by $\sigma \to \infty$. We analyse two possible
cases for these scenarios as regards $\beta$. The first case is for
$\beta$ constant. The second case is for varying $\beta$, in an
adiabatic expansion.

\subsubsection{Constant Temperature}

\begin{figure}[h]
\psfrag{omega}[l]{\hspace*{0.5cm} $\omega$}
\psfrag{sigma}{$\sigma (10^{-16})$}
\psfrag{texto4}{\bf  $\omega_{\theta}(\sigma,\beta)$}
\psfrag{texto1}{\bf $\omega_{\beta}(\sigma,\theta)$}
\psfrag{inflation}{}
\psfrag{betafrio}{}
\psfrag{betaquente}{}
\psfrag{betaquente}{}
\psfrag{thetaalto}{}
\psfrag{thetabaixo}{}
\hspace{-1cm}
{\includegraphics[{height=7.0cm,width=8.0cm}] {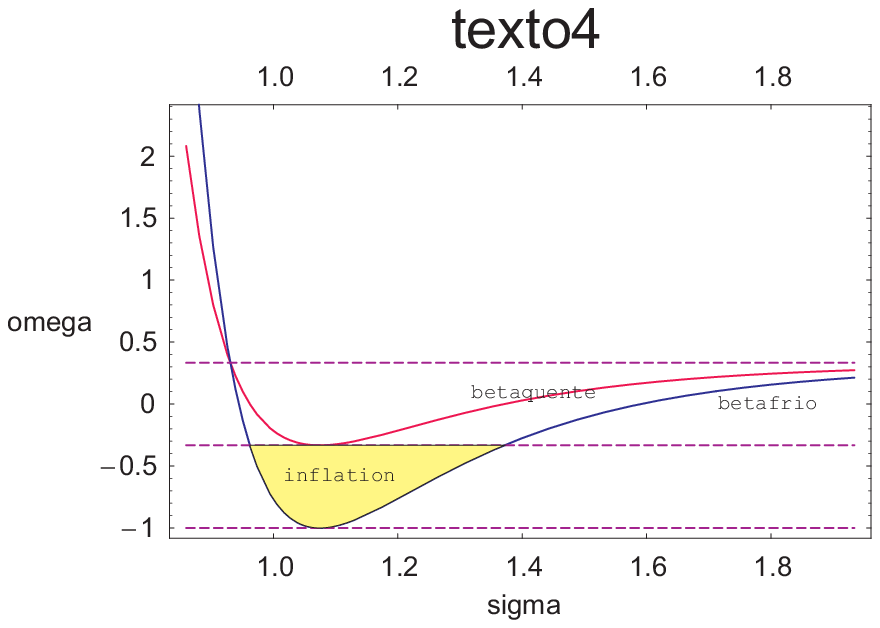}}
{\includegraphics[{height=7.0cm,width=8.0cm}] {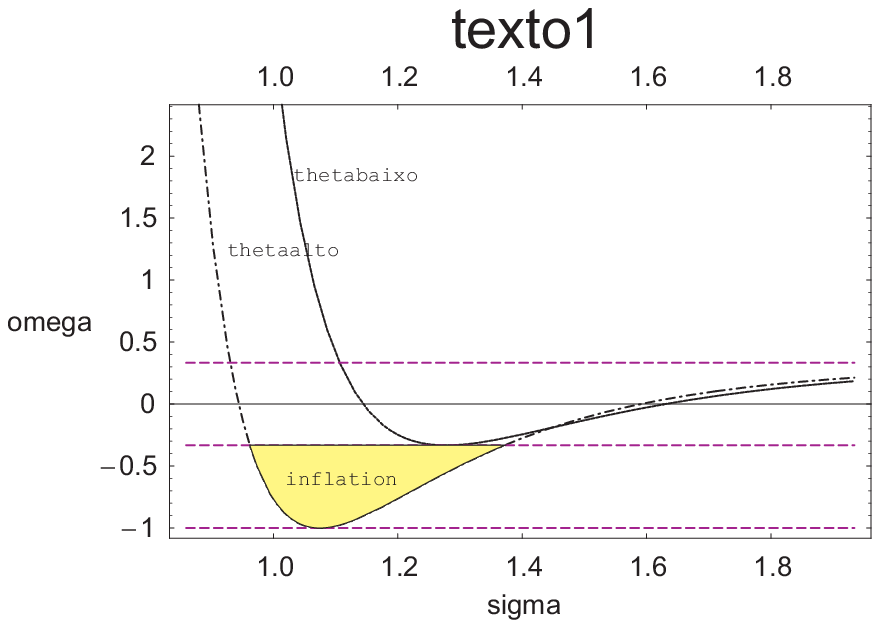}}
\caption{Dashed horizontal lines are $\omega=-1,\;-1/3,\;1/3$. Left Panel: $\theta=2.2\; 10^{-14}$, and values of $\beta$ are chosen to locate the minimum of $\omega$ on the lines $-1$ and $-1/3$. The lower line is plotted for $\beta= \frac{3^{\frac{1}{4}}\theta}{64 \;\sqrt{2}\pi}=10^{-16}$, for the upper line  $\beta= \frac{3^{\frac{1}{4}}\theta}{64\; 2^\frac{3}{4}\pi}=0.8\; 10^{-16}$. Right Panel: $\beta=10^{-16}$, and values of $\theta$ are chosen to locate the minimum of $\omega$ on the lines $-1$ and $-1/3$. The lower line is plotted for $\theta= \frac{64 \;\sqrt{2}\pi \beta}{3^{\frac{1}{4}}}=2.2\; 10^{-14}$, for the upper line $\theta= \frac{64 \;2^{\frac{3}{4}}\pi \beta}{3^{\frac{1}{4}}}=1.1\; 10^{-16}$. Minimum is reached for $\sigma_c=\frac{\theta}{32 \pi}$ }\label{fig1}
\end{figure}

Now let's assume $\beta$ to be fixed, so that $\sigma(\rho)\sim
\frac{1}{\rho^{1/6}}$ (see (\ref{eq:energy-density2})). Thus, at
very high energy density in the early Universe the `compounds' of
the noncommutative field gas feel a very large noncommutative
influence due to very small range $\sigma$ among them. As the
Universe expands the energy density $\rho$ decreases and the range
$\sigma$ increases so that the gas passes through the radiation
dominated regime $\omega(\sigma,\theta,\beta)=1/3$. After that, the
gas enters in the accelerated regime ---  see the lower curve
and delimited shaded region depicted in the left panel of the
Fig.~\ref{fig1}. Finally, the gas achieves again the radiation
dominated regime
 as $\sigma$ becomes sufficiently large --- see
Eq.~(\ref{omega}). Furthermore, the noncommutativity becomes
negligible as $\sigma$ increases in this scenario. This is the
expected scenario that should occurs in any model of inflation.
However, there is a fundamental difference in our scenario described
above,  that is the fact we have fixed both the temperature
($1/\beta$) and $\theta$. This scenario clearly does not exhibit
either cooling or reheating phases in the Universe during the
inflationary regime. This last feature is typical of warm inflation
\cite{Berera:1995ie,Berera:2006xq}.

At the critical point (i.e., $\omega=-1$),
$\theta_c=64\sqrt{\frac{2}{3}}\pi\sigma$ the function (\ref{omega})
assumes the minimum
\ben\label{omega_min2}\omega_{min}(\sigma;\beta)=\frac{1}{3}-\frac{64\beta^4}{27\sigma^4}.
\een

\subsubsection{Adiabatic}

\begin{figure}[h]
\psfrag{omega}{\hspace*{0.5cm} $\omega$}
\psfrag{sigma}{$\sigma (10^{-16})$}
\psfrag{texto8}{\hspace{-2cm}$\omega_\theta(\sigma,\beta)$ - Adiabatic Inflation}
\psfrag{texto1}{\hspace{-15mm}$\omega(\sigma,\beta)$ - Adiabatic Inflation}
\hspace{-1cm}
{\includegraphics[{angle=0,height=7.0cm,angle=0,width=8.0cm}]
{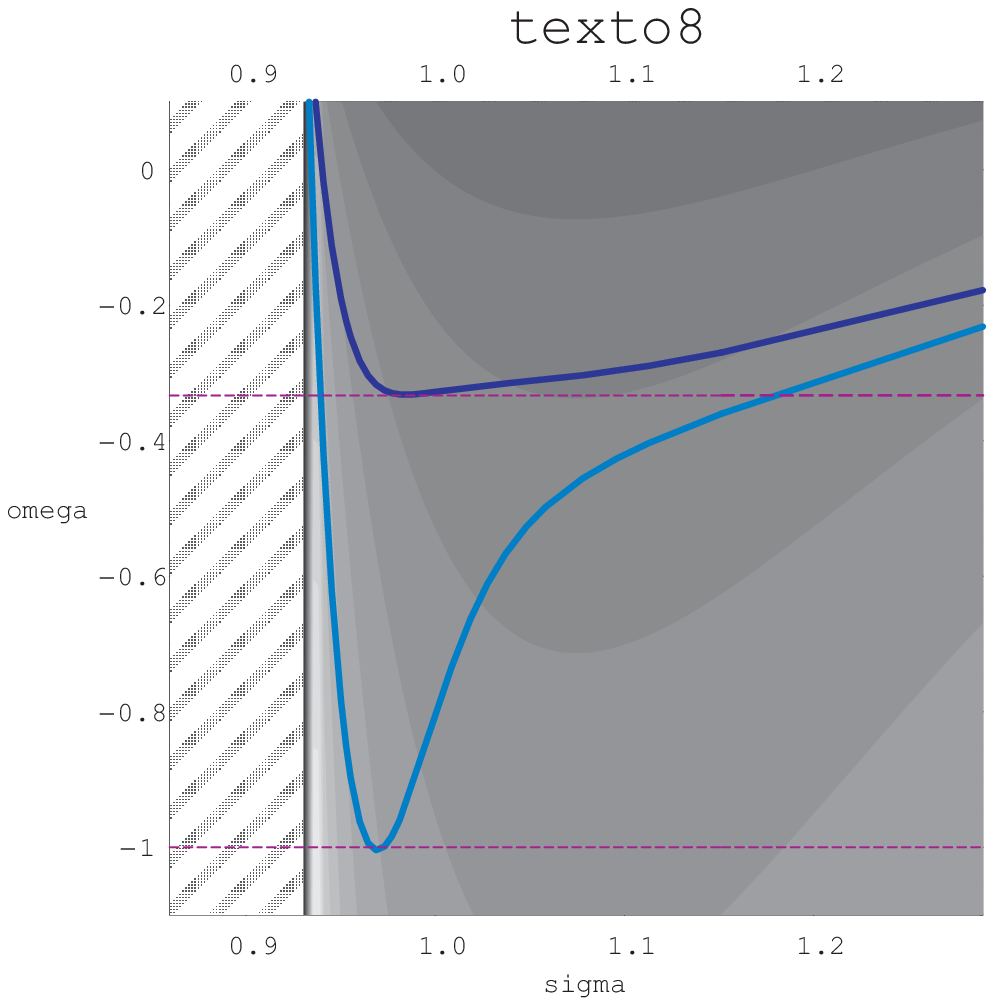}}
{\includegraphics[{angle=0,height=7.0cm,angle=0,width=8.0cm}]
{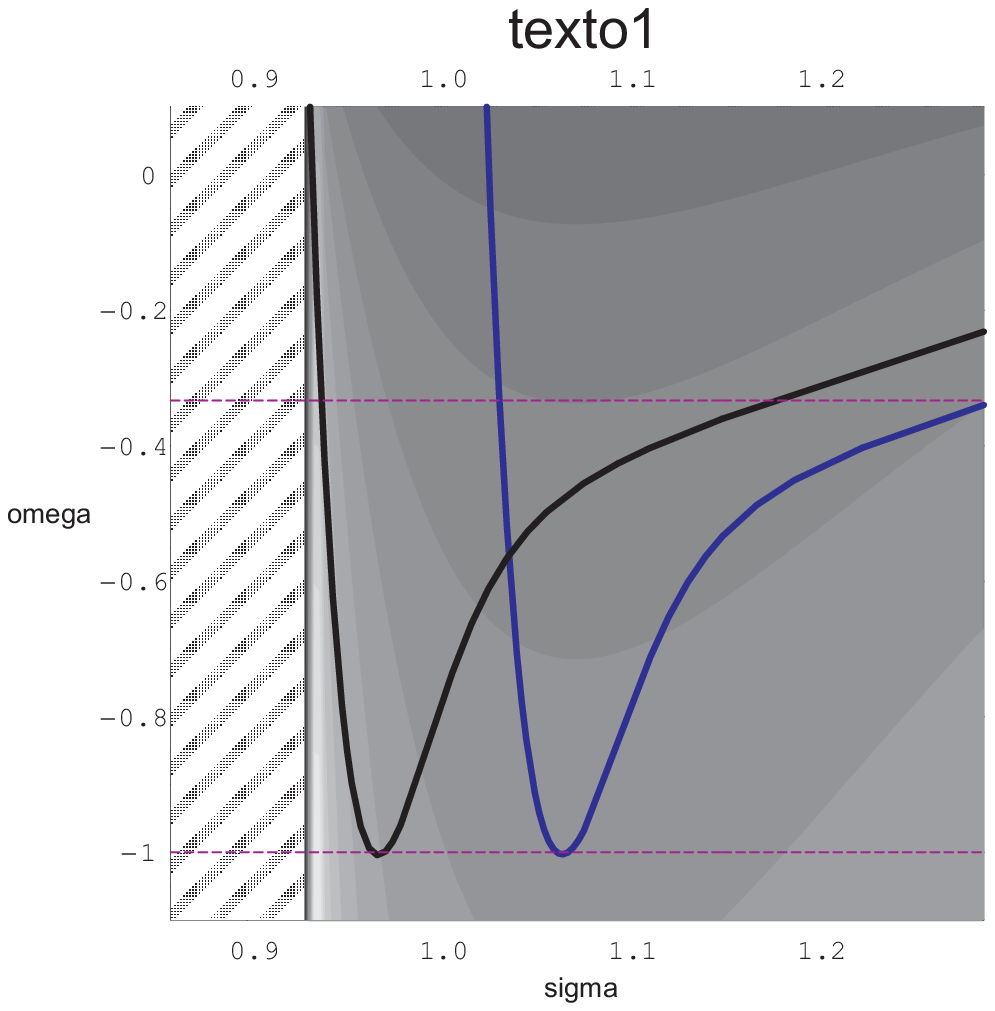}} \caption{Dashed horizontal lines are $\omega=-1,\;-1/3$. Left Panel: $\theta=2.2\; 10^{-14}$, the curves are obtained numerically in such a way that their minima are located at $\omega=-1$ and $\omega=-1/3$ for the lower and upper curves ($s=1.8 \;10^{64}$ and $s=2.1 \;10^{64}$) respectively. Right panel shows that increasing $\theta$ the minimum of $\omega$ is displaced to larger $\sigma$ (later limes). Curves are depicted for $s=1.8\; 10^{64},\; \theta=2.2\; 10^{-14}$ (left) and $s=1.2\; 10^{64},\; \theta=2.4\; 10^{-14}$ (right). Pattern filled regions are not covered by any lines. Shading represents contour lines for constant $\beta$, lighter regions mean greater $\theta$ (and smaller temperature).}\label{fig2}
\end{figure}

Now we assume $\beta$ as variable. In this case we may have very
distinct variations of temperature. In particular, one interesting
possibility we are going to analyse here is the adiabatic expansion
of the Universe. Therefore, in this case the molar entropy
(\ref{eq:entropy}) is constant.

In this scenario, as the Universe expands adiabatically, its
temperature initially increases --- ~see Fig.~\ref{fig2}.

Then depending on the value of $\theta$, it enters in an inflation
regime ($\omega<-1/3$). In that regime the Universe keeps cooling
until it reaches a minimum value for $\omega$. After that it starts
to reheat until it leaves out the inflation regime. Therefore just
after inflation the Universe is hot. One may even observe that some
time after inflation the Universe starts to cool again. Also at some
point the noncommutativity will be negligible, since $\sigma$ is
increasing and $\theta$ is fixed. The most interesting feature of
this scenario is that the Universe shows up a cooling and then
reheating phase as $\sigma$ increases (or equivalently, $\rho$
decreases).

However, this scenario is the most difficult to analyze analytically
and we had to rely on numerical calculations.

A common feature of both above scenarios is that increasing $\theta$
moves the minimum of the $\omega$ function to greater values of
$\sigma$ and consequently later times. Since inflation time is
deeply connected to the position of this minimum, locating the
inflation era fixes the value of $\theta$, which can be
cross-checked with other empirical bounds for the strength of the
noncommutativity.

\subsection{Inflationary Scenarios with Fixed $\sigma$}

Now we consider the noncommutativity range parameter fixed,
inherently tied to the short distance physics at all times and
consider the possible inflationary scenarios, by constraining
$\beta$ and $\theta$. In these scenarios, the classical radiation
dominated era may be achieved by $\theta \to 0$.

\subsubsection{Constant Temperature}

\begin{figure}[h]
\psfrag{omega}{\hspace*{0.5cm} $\omega$}
\psfrag{theta14}{$\theta\; (10^{-14})$}
\psfrag{texto5sigmafixo}{\hspace{1cm}$\omega_\sigma(\theta,\beta)$}
\psfrag{texto6betafixo}{\hspace{1cm}$\omega_\beta(\theta,\sigma)$}
\hspace{-1cm}
{\includegraphics[{angle=0,height=7.0cm,angle=0,width=8.0cm}]
{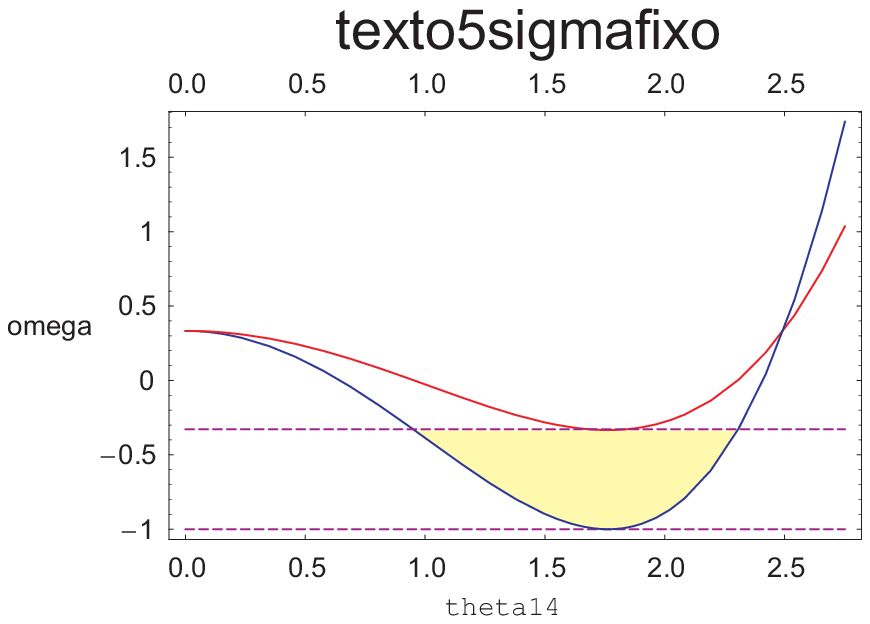}}
{\includegraphics[{angle=0,height=7.0cm,angle=0,width=8.0cm}]
{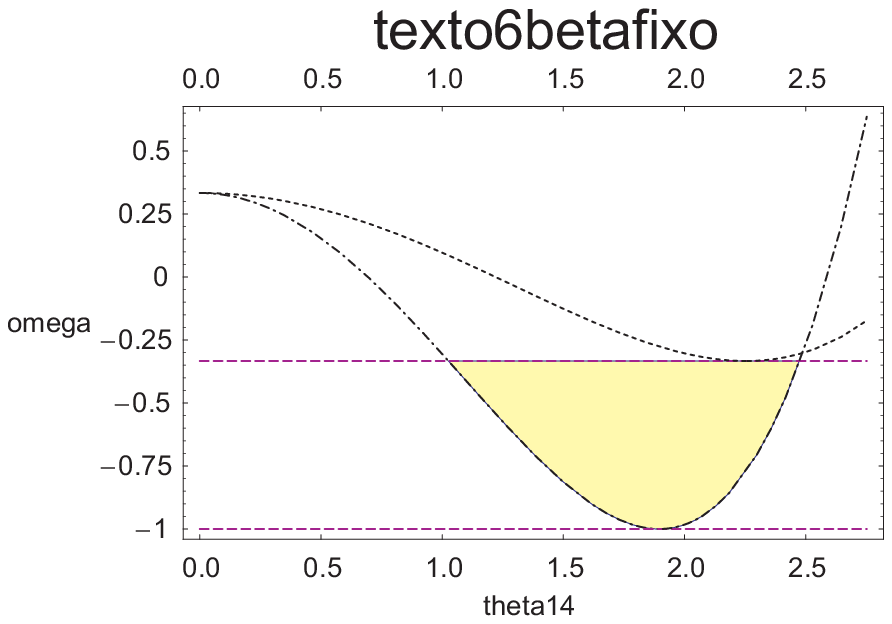}} \caption{The function $\omega(\theta)$. In the
right panel the temperature is fixed, $\beta\!=\!10^{-16}$, for both
curves. The curves are depicted for $\sigma_{\rm min}=\frac{2\,
\beta}{\sqrt{3}}$ (for $\omega=-1$) and $\sigma_{\rm max}=\frac{2\,
2^{1/4}\beta}{\sqrt{3}}$ (for $\omega=-1/3$). In the left panel the
parameter $\sigma$ is fixed, $\sigma=\sigma_{c}$, for both curves.
The curves are depicted for $\beta_{\rm
min}=\frac{\sigma_c\sqrt{3}}{2}$ and $\beta_{\rm
max}=\frac{\sqrt{3}\sigma_c}{2\, 2^{1/4}}$, where $\sigma_c$ is
defined as $\theta_c=64\sqrt{\frac{2}{3}}\pi\sigma_c$, being
$\theta_c$ the point that minimizes $\omega$.}\label{fig3}
\end{figure}

Let's now assume again $\beta$ to be fixed, so that
$\theta(\rho)\sim\rho^{1/2}\sigma^3$ (see
(\ref{eq:energy-density2})). In the regime of $\rho\to0$, we find
that $\theta\to 0$ in a faster way. Now by using the original
Eq.~(\ref{omega}) we find
$\omega(\sigma,\theta,\beta)\to\frac{1}{3}$, which corresponds again
to a radiation dominated regime --- see Fig.~\ref{fig3}, the left
panel.

\begin{figure}[h]
\psfrag{omega}{\hspace*{0.5cm} $\omega$}
\psfrag{beta}{$\beta\; (10^{-16}))$}
\psfrag{textobeta}{$\omega_\theta(\beta,\sigma)$}
\psfrag{textobetasigma}{\hspace{1cm}$\omega_\sigma(\beta,\theta)$}
\hspace{-1cm}
{\includegraphics[{angle=0,height=7.0cm,angle=0,width=8.0cm}]
{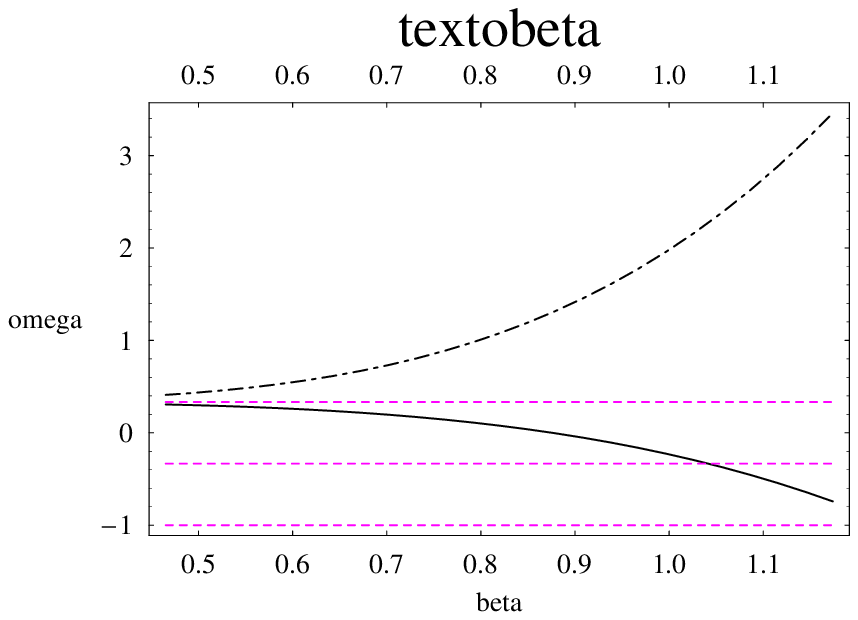}}
{\includegraphics[{angle=0,height=7.0cm,angle=0,width=8.0cm}]
{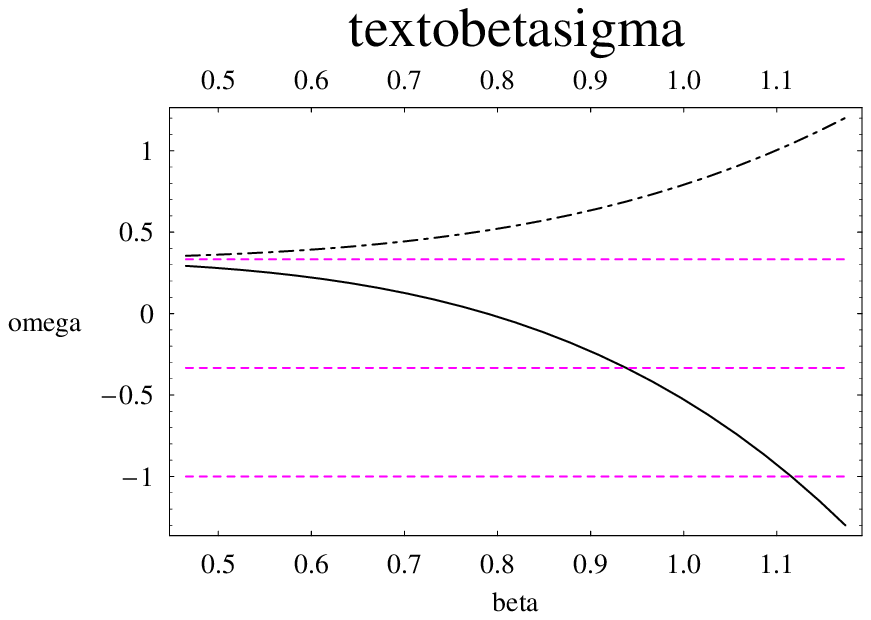}} \caption{The function $\omega(\beta)$. In the
left panel the parameter $\theta$ is fixed, $\theta\!=1.8\; 10^{-14}$, for
both curves. {\ The curves are depicted for $\sigma_{\rm 1}=1.4\; 10^{-16}$
(inflation) and $\sigma_{\rm 2}=1.8 \; 10^{-16}$ (no-inflation). In the right
panel the parameter $\sigma$ is fixed, $\sigma=1.1\; 10^{-16}$, for both
curves. The curves are depicted for $\theta_{\rm 1}=2.6\; 10^{-14}$ (inflation)
and $\theta_{\rm 2}=2.3\; 10^{-14}$ (no-inflation)}.}\label{fig4}
\end{figure}

Let's analyse what happens at the critical value (at $\omega=-1$) of
$\sigma$, i.e., $\sigma_c=\frac{\theta}{64\pi}$. At the
critical point, $\sigma_c=\frac{\theta}{64\pi}$ the function
(\ref{omega}) assumes the minimum
\ben\label{omega_min}\omega_{min}(\theta,\beta)=\frac{1}{3}-\frac{2^{28}\pi^4\beta^4}{9\theta^4}.
\een At low temperature, the minimum of the function
$\omega(\theta;\beta)$ again tends to cross $\omega=-1$ --- see
Fig.~\ref{fig4}, the right panel.

Replacing this critical value into (\ref{eq:energy-density2}), we
may obtain a relation between $\theta$ and the energy density as
\ben\theta\sim\frac{1}{\rho^{1/4}}. \een As the Universe expands
$\rho\to0$ and then $\theta\to\infty$. As consequence of this
$\omega_{min}(\theta,\beta)\to\frac{1}{3}$, which comprises again a
radiation dominated regime.

\section{Conclusions}
\label{conclu}

In this paper we have used noncommutative fields to find
cosmological scenarios that can develop inflation at early Universe.
A gas composed with a noncommutative bosonic field shows up several
interesting properties. In particular, in the dilute regime at high
temperature, these properties can be read off from Eq.
(\ref{omega}).

The aim of the paper was essentially to find the
equation of state of the gas by tackling approximately the
statistical quantum mechanics problem for the noncommutative field
gas at high temperature. Since we are dealing with highly relativistic
bosons not necessarily massless, the approximate relation between
energy and momenta $E\simeq p$ used along the paper is valid. This
in turn is consistent with the formulation of free massless boson
with noncommutative target space adopted previously. The grand
canonical ensemble used in the quantum statistical formulation is also
valid because, on the other hand, the masslessness of the bosons is
just an approximation.

The noncommutative field gas we are dealing with here can develop
inflation for a period of time that can be adjusted to observational
data. We showed some scenarios that exit inflation phase to a
radiation dominated epoch, as it should be according to the
perspective of the inflationary cosmology. This feature is certainly
a motivation for further studies of these scenarios. In particular,
the scenarios with $\theta$ fixed.

In scenarios with $\sigma$ and $\beta$ fixed, we are left with
only the parameter $\theta$ free to be determined for a given
temperature in Eq. (\ref{omega_min}). Let us estimate the value of
$\theta$ for $T=10^{14}GeV\simeq10^{27} K$, the approximate
temperature at beginning of the inflation, for $\omega_{min}\geq-1$
and $\omega_{min}<-1/3$, respectively \ben \theta_{-1}\geq 0.4\times
10^{-25}cm, \qquad \theta_{-1/3}<0.5\times 10^{-25}cm. \een This
result should be contrasted with other astrophysical bounds on the
parameter $\theta.$ For instance, in \cite{Carmona:2002iv}, it was
estimated a value for $\theta$ to be of the order of $(10^{20}
eV)^{-1}\simeq10^{-25}cm$. They have used the fact that the highest
cosmic ray energy be no greater than $\sim 10^{20} eV$. Recall that
the GZK cutoff is of the order of $10^{19} eV$. Furthermore, as we can easily see above, there is
a short range for $\theta$ that allows for inflation in these scenarios.

A proper understanding of the nature of the parameter $\sigma$, for
instance, a dynamical theory for it, could shed more light on these
developments. Further possibilities may arise if one consider other
kinds of regularization for the delta function in (\ref{eqNCTA:69}).

Several other interesting issues such as the implications of the
sound speed and the density perturbations can also be investigated
to know in what extent the model is in accordance with CMB data. One
can estimate the adiabatic sound speed by using \ben
c_s^2=\left(\frac{\partial P}{\partial
\rho}\right)_{S,N}=-\frac{V^2}{m_0N}\left(\frac{\partial P}{\partial
V}\right)_{S,N}, \een with $\rho=m/V=m_0N/V$. Once we have
previously assumed a dilute gas $(z\ll1)$ at high temperature, then
by using eqs.(\ref{therm_quantity2})-(\ref{log_grand_approx}) one
can easily show that the gas satisfies $PV=NT$. The sound speed is
then given by \ben c_s=\sqrt{\frac{T}{m_0}}, \een where $m_0$ is the
mass of ultra relativistic bosons satisfying $E^2=m_0^2+p^2\simeq
p^2$. So, at first order in $z$, the approximation maintained
throughout this work, the sound speed does not depend on
non-commutative parameters. The first correction comes from higher
orders in $z$ expansion which can be computed quite easily to be:
\begin{equation}
\frac{T\,\left( 1 - \frac{7\,z}{64} \right) }{m_0} +
\frac{91\,z\,{\theta }^2}{16384\,m_0\,{\pi }^2\,T^3\,{\sigma }^6} +
  \frac{z\,\log (z)}{64\,m_0\,T}.
\end{equation}

Since in this work we consider the high temperature limit, in which
we can disregard the mass, further analysis on the speed of sound
cannot be pursued. We let further studies related to this issue for
future investigations, together with the low temperature limit.

 \acknowledgments

This work was partially supported by Conselho Nacional de
Desenvolvimento Cient\'\i fico e Tecnol\'ogico (CNPq).



\end{document}